\documentclass[prl,aps,twocolumn]{revtex4}

\usepackage{amsmath,graphicx,epsfig}
\usepackage{epstopdf}
\usepackage{euscript}
\usepackage{amsfonts}
\usepackage{amssymb}
\usepackage{float}

\def\prn#1{{\left(#1\right)}}

\def\bra#1{{\langle#1|}}

\def\cg(#1,#2)(#3,#4)(#5,#6){\bra{#1,#2,#3,#4}#5,#6\rangle}

\def\ts#1{{_{\mbox{\scriptsize #1}}}}

\def\threej(#1,#2)(#3,#4)(#5,#6){\begin{pmatrix}#1&#3&#5\\#2&#4&#6\end{pmatrix}}
\def\sixj(#1,#2,#3)(#4,#5,#6){\begin{Bmatrix}#1&#2&#3\\#4&#5&#6\end{Bmatrix}}
\def\ninej(#1,#2,#3)(#4,#5,#6)(#7,#8,#9){\begin{Bmatrix}#1&#2&#3\\#4&#5&#6\\#7&#8&#9\end{Bmatrix}}

\def\sR{{\ensuremath{\EuScript R}}}

\def\sV{{\ensuremath{\EuScript V}}}

\def\mb{\mathbf}
\def\bs{\boldsymbol}

\newlength{\defbaselineskip}
\newcommand{\setlinespacing}[1]%
           {\setlength{\baselineskip}{#1 \defbaselineskip}}

\begin{document}

\title{Constraints on long-range spin-gravity and monopole-dipole couplings of the proton} 

\author{Derek F. Jackson Kimball}
\email{derek.jacksonkimball@csueastbay.edu}
\affiliation{Department of Physics, California State University --
East Bay, Hayward, California 94542-3084, USA}

\author{Jordan Dudley}
\affiliation{Department of Physics, California State University --
East Bay, Hayward, California 94542-3084, USA}

\author{Yan Li}
\affiliation{Department of Physics, California State University --
East Bay, Hayward, California 94542-3084, USA}

\author{Dilan Patel}
\affiliation{Department of Physics, California State University --
East Bay, Hayward, California 94542-3084, USA}

\author{Julian Valdez}
\affiliation{Department of Physics, California State University --
East Bay, Hayward, California 94542-3084, USA}

\date{\today}



\begin{abstract}
Results of a search for a long-range monopole-dipole coupling between the mass of the Earth and rubidium (Rb) nuclear spins are reported.  The experiment simultaneously measures the spin precession frequencies of overlapping ensembles of $^{85}$Rb and $^{87}$Rb atoms contained within an evacuated, antirelaxation-coated vapor cell.  The nuclear structure of the Rb isotopes makes the experiment particularly sensitive to spin-dependent interactions of the proton. The spin-dependent component of the gravitational energy of the proton in the Earth's field is found to be smaller than $3 \times 10^{-18}~{\rm eV}$, improving laboratory constraints on long-range monopole-dipole interactions by over three orders of magnitude.
\end{abstract}



\maketitle

The standard model of particle physics and general relativity provide frameworks for understanding a vast array of phenomena. Nonetheless, there remain important observations that these foundational theories cannot explain, such as the nature of dark matter, the asymmetry between matter and antimatter, and the accelerating expansion of the universe.  Such unexplained mysteries motivate searches for new fundamental forces, fields, and particles.  A heretofore undiscovered spin-0 field composed, for example, of axion-like particles (ALPs), could be the dark matter observed throughout the universe \cite{Din83,Pre83,Duf09}. A scalar-pseudoscalar coupling of such a field to matter violates parity (P) and time-reversal (T) symmetries \cite{Moo84}, and thus might be connected to the observed matter-antimatter asymmetry of the universe \cite{Moh06}. Furthermore, a massless or nearly massless spin-0 field can manifest as a dark energy over cosmological distances causing acceleration of the universe's expansion \cite{Rat88,Wet88,Cal98}. Spin-0 fields appear naturally in extensions of the Standard Model such as string theory \cite{Svr06,Arv10}, in possible solutions to the hierarchy problem \cite{Gra15}, and generically in theories featuring spontaneous symmetry breaking \cite{Pec77,Wei78,Wil78,Kim79,Shi80,Din81,Wil82,Gel83,Gel81,Chi81,Ans82}. Low-mass, spin-0 fields are ubiquitous features of theoretical attempts to address the most important problems in modern physics.

Of particular interest for laboratory tests is the fact that light spin-0 fields with pseudoscalar couplings to matter lead to long-range spin-dependent potentials \cite{Moo84,Dob06,Fla09}.  If the new field is considered to be an additional component of gravity, as suggested by certain scalar-tensor extensions of general relativity based on a Riemann-Cartan spacetime \cite{Heh76,Sha02,Ham02,Kos08}, there would be coupling of spins to gravitational fields, causing particles to acquire a gravitational dipole moment (GDM).  The dominant gravitational field in a laboratory setting is that due to the Earth, which generates a spin-dependent Hamiltonian with the nonrelativistic form \cite{Lei64,Har76,Per78,Fla09}:
\begin{align}
H_g = k_i \frac{\hbar}{c} \bs{\sigma}_i\cdot\bs{g} = \chi_i \bs{\sigma}_i\cdot\bs{g} = \hbar\Omega_{gi}
\label{Eq:GDM-Hamiltonian}
\end{align}
where $k_i$ is a dimensionless parameter setting the scale of the new interaction for particle $i$, $\bs{\sigma}_i$ is the intrinsic spin of particle $i$ in units of $\hbar$, $\bs{g}$ is the acceleration due to gravity, $\chi_i=k_i\hbar/c$ is the particle's gyro-gravitational ratio, and $\Omega_{gi}$ is the particle's spin precession frequency due to Earth's gravitational field.  If the strength of the pseudoscalar coupling is the same as that of the ordinary (tensor) gravitational coupling, $k_i \approx 1$ \cite{Fla09}.  Another framework for analyzing such exotic spin-mass couplings, known as the Moody-Wilczek-Dobrescu-Mocioiu (MWDM) formalism \cite{Moo84,Dob06}, assumes one-boson exchange within a Lorentz-invariant quantum field theory, in which case a light pseudoscalar field generates a monopole-dipole potential $\sV_{9,10}(r)$ of the form (the subscript refers to the MWDM potentials enumerated in Ref.~\cite{Dob06}):
\begin{align}
\sV_{9,10}(r) = \frac{{\rm g}_p^i{\rm g}_s^j\hbar}{8\pi m_ic} \bs{\sigma}_i \cdot \hat{\mb{r}} \prn{ \frac{1}{r\lambda} + \frac{1}{r^2} }  e^{-r/\lambda} ~,
\label{Eq:monopole-dipole-potential}
\end{align}
where ${\rm g}_p^i$ is the pseudoscalar coupling constant to particle $i$, ${\rm g}_s^j$ is the scalar coupling constant to particle $j$, $m_i$ is the mass of particle $i$, $\mb{r}=r\hat{\mb{r}}$ is the displacement vector between $i$ and $j$, and $\lambda$ is the range of the new force. Equations \eqref{Eq:GDM-Hamiltonian} and \eqref{Eq:monopole-dipole-potential} can be connected in the limit $\lambda \gg R_E$, where $R_E$ is the radius of the Earth, by integrating the contribution of the constituent particles making up the Earth, assuming ${\rm g}_s$ is roughly equal for protons and neutrons (and neglecting the scalar coupling to electrons, which is generally treated separately):
\begin{align}
k_i =  \frac{c}{\hbar} \chi_i \approx \frac{1}{8\pi}\frac{{\rm g}_p^i{\rm g}_s}{\hbar c} \frac{M\ts{Pl}^2}{m_p m_i}~,
\label{Eq:k}
\end{align}
where $M\ts{Pl} = \sqrt{ \hbar c / G }$ is the Planck mass.

The most stringent constraints on GDMs have been established by previous experiments using an electron-spin-polarized torsion pendulum \cite{Hec08} and measuring spin-precession of $^{199}$Hg and $^{201}$Hg \cite{Ven92} (Table~\ref{Table:k-limits}).  These previous experiments searched for electron or neutron GDMs \cite{Kim15}. In contrast, our experiment is sensitive to the proton GDM, and improves upon the existing laboratory constraint \cite{You96} by over three orders of magnitude. There are a number of plausible theoretical models in which exotic monopole-dipole couplings to neutron spins are strongly suppressed relative to those for proton spins \cite{Raf99,Raf12}, and so it is sensible to regard the neutron and proton GDM constraints independently.

\begin{table}
\caption{Constraints (at the 90\% confidence level) on the dimensionless spin-gravity coupling parameter $k$ [Eq.~\eqref{Eq:GDM-Hamiltonian}].}
\medskip \begin{tabular}{lcl} \hline \hline
Particle~~~~~~~ & Upper limit on $k$~~~~~~~ & Experiment~~ \\
\hline
\rule{0ex}{3.6ex} Electron & $10$ & Ref.~\cite{Hec08} \\
\rule{0ex}{3.6ex} Neutron & $10^3$ & Ref.~\cite{Ven92} \\
\rule{0ex}{3.6ex} Proton & $3 \times 10^8$ & Ref.~\cite{You96} \\
\rule{0ex}{3.6ex} Proton & $2 \times 10^5$ & This work \\
\hline \hline
\end{tabular}
\label{Table:k-limits}
\end{table}

\begin{figure}
\includegraphics[width=3.7 in]{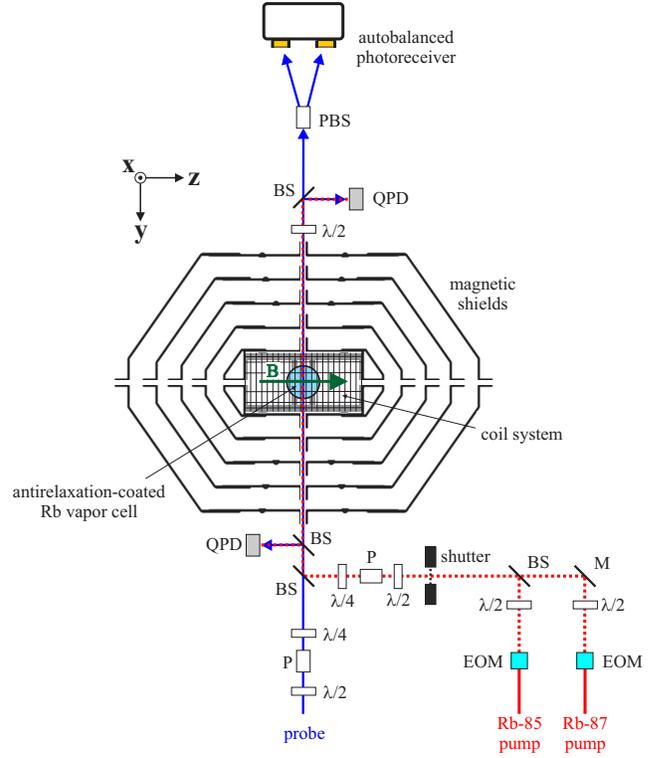}
\caption{Schematic of the experimental setup. P~=~linear polarizer, M~=~mirror, BS~=~(nonpolarizing) beamsplitter, PBS~=~polarizing beamsplitter, $\lambda/4$~=~quarter-wave plate, $\lambda/2$~=~half-wave plate, EOM~=~electro-optic modulator, QPD~=~quadrant photodiode. Designation of $\mb{x}$, $\mb{y}$, and $\mb{z}$ directions is shown in the upper left corner. Red solid and dashed lines represent the pump beams, blue arrows represent the probe beam. The green arrow at the center of the diagram represents the applied magnetic field $\mb{B}$, which is directed along $\bs{\Omega}_E$ (see Fig.~\ref{Fig:ExperimentalGeometry}). Assorted optics and electronics for laser control, data acquisition, and experiment control are not pictured.}
\label{Fig:ExperimentalSetup}
\end{figure}

\begin{figure}
\includegraphics[width=3.0 in]{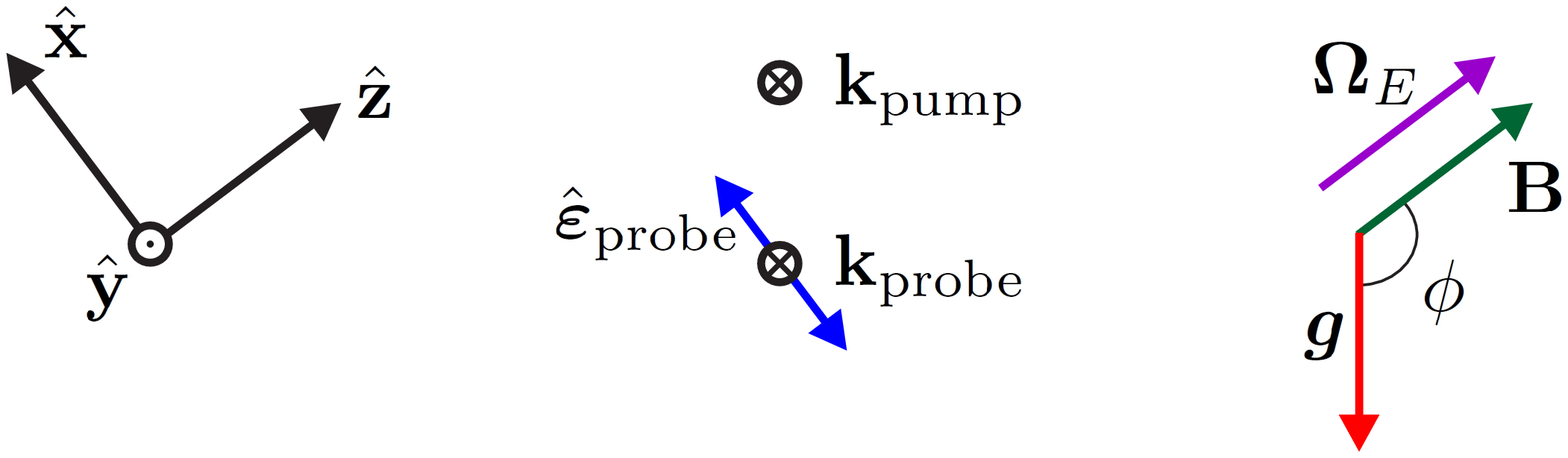}
\caption{Diagram showing the experimental geometry. The Earth's angular velocity vector $\bs{\Omega}_E$ (purple arrow) is along $\hat{\mb{z}}$, $\bs{g}$ is the local gravitational field of the Earth (red arrow), which is at an angle $\phi \approx 128^\circ$ to $\bs{\Omega}_E$. The magnetic field $\mb{B}$ (green arrow) is applied along $\pm \hat{\mb{z}}$ (here pictured along $+\hat{\mb{z}}$). The pump and probe beams are collinear: $\mb{k}\ts{pump}$ and $\mb{k}\ts{probe}$ are the pump and probe beam wave vectors, respectively, both along $-\hat{\mb{y}}$. The probe beam has linear polarization $\hat{\bs{\varepsilon}}\ts{probe}$ (blue double-arrow) along the $x$-axis.}
\label{Fig:ExperimentalGeometry}
\end{figure}

In the present work, we have used a dual-isotope rubidium (Rb) comagnetometer \cite{Kim13} to search for a coupling between Rb nuclear spins and the Earth's gravitational field [Eq.~\eqref{Eq:GDM-Hamiltonian}] or to the mass of the Earth via a long-range monopole-dipole interaction [Eq.~\eqref{Eq:monopole-dipole-potential}].  The basic concept of our experiment is to use synchronous laser optical pumping to generate precessing spin polarization of Rb atoms transverse to a uniform magnetic field $\mb{B}$, and then employ off-resonant laser light to simultaneously measure the spin precession frequencies of $^{85}$Rb and $^{87}$Rb (Fig.~\ref{Fig:ExperimentalSetup}). The field $\mb{B}$ is directed along the Earth's angular velocity $\bs{\Omega}_E$ (Fig.~\ref{Fig:ExperimentalGeometry}) in order to minimize systematic error due to the gyro-compass effect, which is related to the fact that the laboratory is a noninertial reference frame due to Earth's rotation \cite{Ven92,Hec08,Bro10,Tul13}. Considering only Larmor precession, the gyro-compass effect, and a possible spin-gravity coupling, the $^{85}$Rb and $^{87}$Rb spin-precession frequencies, as discussed in Refs.~\cite{Kim13,Kim15}, are given by
\begin{align}
\Omega_{85}\prn{\pm} & \approx \frac{\gamma_{85}B}{\hbar} \pm \Omega_E \pm \prn{\frac{1}{6}\chi_e - \frac{5}{42} \chi_p} \frac{g\cos\phi}{\hbar}~, \\
\Omega_{87}\prn{\pm} & \approx \frac{\gamma_{87}B}{\hbar} \pm \Omega_E \pm \prn{\frac{1}{4}\chi_e + \frac{1}{4} \chi_p} \frac{g\cos\phi}{\hbar}~,
\label{Eq:Rb85-Rb87-precession-frequencies}
\end{align}
where $\phi \approx 128^\circ$ is the angle between $\bs{\Omega}_E$ and $\bs{g}$ (Fig.~\ref{Fig:ExperimentalGeometry}), $\gamma_{85}$ and $\gamma_{87}$ are the gyromagnetic ratios ($\gamma = g_F\mu_0$, where $g_F$ is the Land\'e factor for the hyperfine level with total angular momentum $F$ and $\mu_0$ is the Bohr magneton), and $\pm$ refers to the cases where $\mb{B}$ is directed parallel and antiparallel to $\bs{\Omega}_E$ (i.e., the $+$ case corresponds to $\mb{B}$ pointing toward the North Star). More specifically, $\Omega_{85}$ is the precession frequency of $^{85}$Rb atoms in the $F=3$ ground state hyperfine level and $\Omega_{87}$ is the precession frequency of $^{87}$Rb atoms in the $F=2$ ground state hyperfine level. To analyze the data, we construct the ratio
\begin{align}
\sR_{\pm} = \frac{\Omega_{87}\prn{\pm} - \Omega_{85}\prn{\pm}}{\Omega_{87}\prn{\pm}+\Omega_{85}\prn{\pm}}~.
\label{Eq:R=SpinPrecessionRatio}
\end{align}
Accounting for the fact that $\gamma B \gg \chi_e g,~\chi_p g,~\hbar\Omega_E$ and neglecting nuclear magnetic moments,
\begin{align}
\Delta \sR &= \sR_{+} - \sR_{-} \approx \prn{\frac{\gamma_{87}-\gamma_{85}}{\gamma_{87}+\gamma_{85}}} \prn{\frac{4\chi_p g \cos\phi + 10\hbar\Omega_E}{\mu_0 B}}.
\label{Eq:DeltaR}
\end{align}
$\Delta \sR$ is sensitive to the proton GDM while there is first-order cancellation of effects related to Larmor precession and an electron GDM \cite{Kim13}.

\begin{figure}
\includegraphics[width=3.0 in]{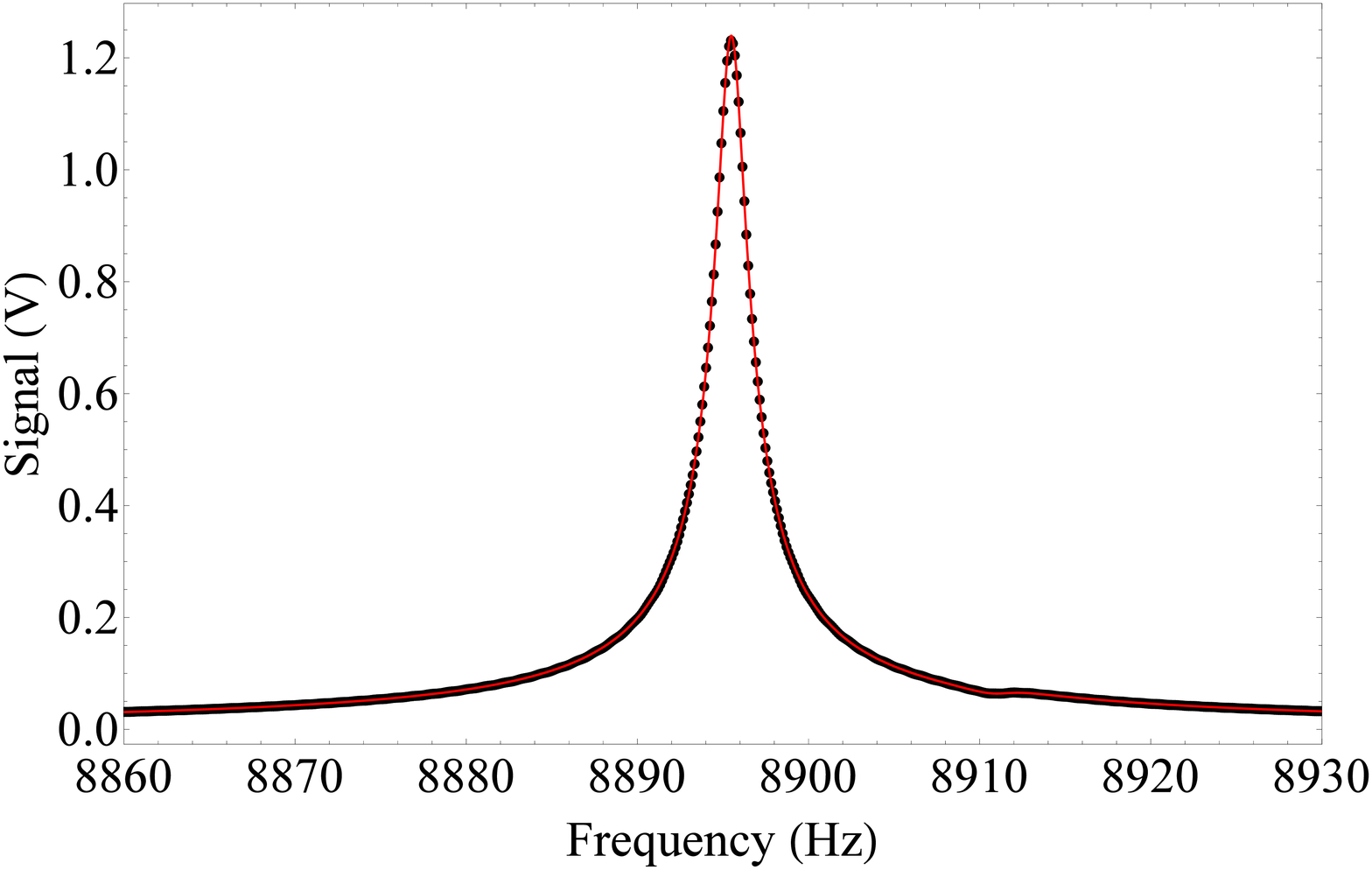}
\includegraphics[width=3.0 in]{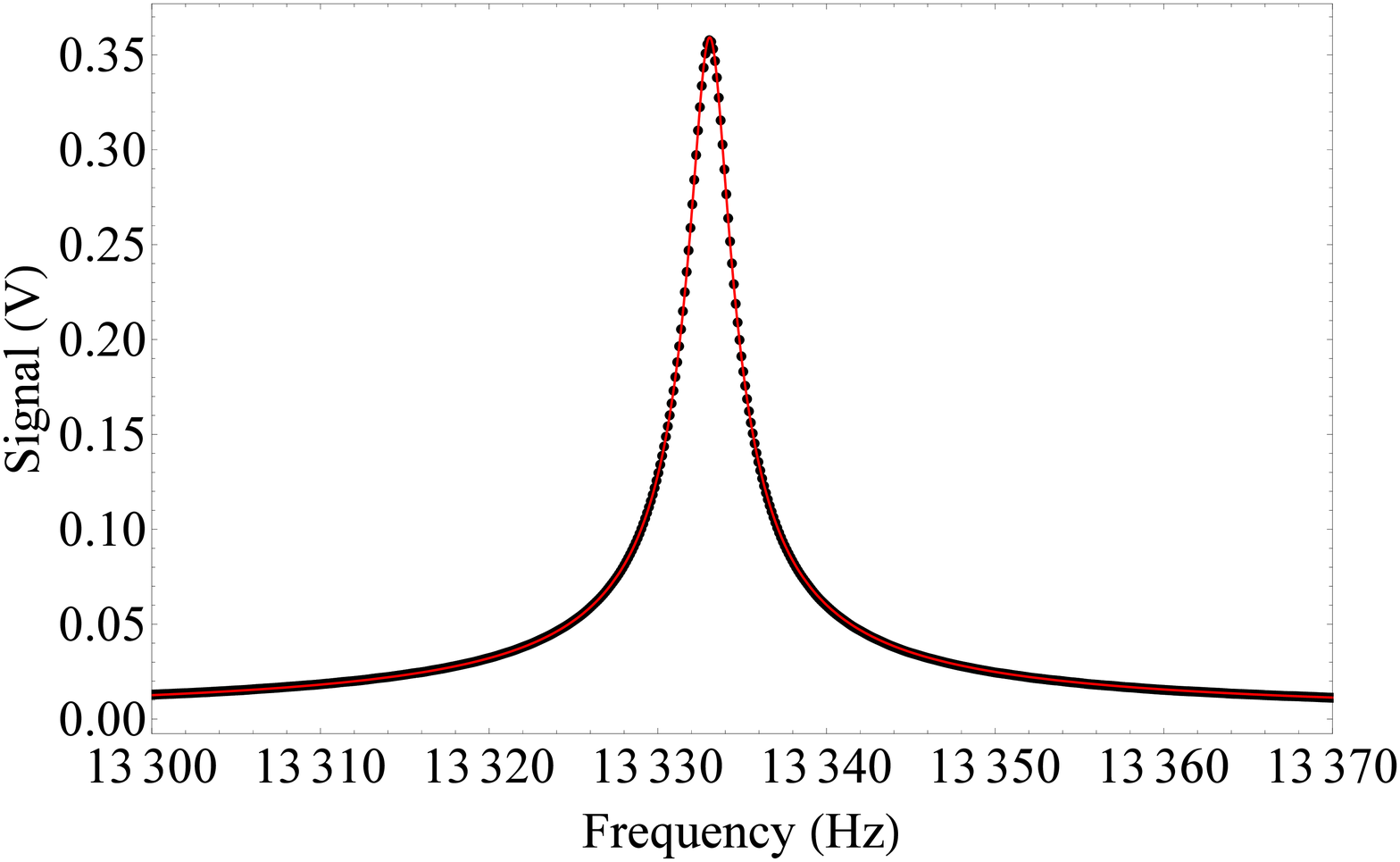}
\caption{Example of a single $\approx 1~{\rm s}$ data sample from the pump/probe spin precession measurement in the frequency domain, showing the Fourier transform of the probe optical rotation data (black dots) and Lorentzian fits (red lines) which determine $\Omega_{85}$ and $\Omega_{87}$. The upper plot is centered around $\Omega_{85}$ and the lower plot is centered around $\Omega_{87}$. A small amplitude feature in the upper plot is observed at $\approx 8911~{\rm Hz}$; this is optical rotation from a far off-resonant transition related to precession of spins in the $^{85}$Rb $F=2$ ground state hyperfine level and is accounted for in the fit. In order to obtain reliable fits, $B$ must be sufficiently large so that this small amplitude feature is well-separated from the main peak, see Ref.~\cite{Kim13} for further details. For this data, the alkene-coated vapor cell was used and $B = 19.052729(3)~{\rm mG}$.}
\label{Fig:SampleDataFrequency}
\end{figure}

At the heart of the experimental setup is a Rb vapor (natural isotopic mixture: $\approx 72\%$ $^{85}$Rb, $\approx 28\%$ $^{87}$Rb) contained in an evacuated, spherical (5-cm diameter), antirelaxation-coated glass cell. Both alkene-coated \cite{Bal10} and paraffin-coated \cite{Ale02} cells were used to check for cell- and coating-related systematic errors during the experiments. The cell is located inside a set of nine independent magnetic field coils that enable control of longitudinal and transverse components of $\mb{B}$ as well as all first-order gradients and the second-order gradient along $\mb{B}$. The cell and coil system are nested within a five-layer mu-metal shield system that provides near uniform shielding of external fields to a part in $10^7$ \cite{Xu06,Kim16}. The outermost shield layer is temperature stabilized using a resistive heater and the shield layers are spaced with foam that provides thermal insulation and acoustic damping; the vapor cell temperature is $\approx 28.5^\circ{\rm C}$ yielding a Rb vapor density of $\approx 2\times 10^{10}~{\rm atoms/cm^3}$. Temperature stabilization of the shields significantly reduces thermal drift of the magnetic field conditions within the innermost shield layer. (Note that the effect of an exotic spin-dependent interaction, either gravity or a long-range monopole-dipole coupling, is not screened by the magnetic shield as discussed in Ref.~\cite{Kim16}.)

Measurement of $\Omega_{85}$ and $\Omega_{87}$ is carried out using a temporally separated pump/probe sequence. During the $\approx 1~{\rm s}$ synchronous optical pumping stage, the Rb atoms are illuminated by two collinear, $\approx 2~{\rm mm}$ diameter, circularly polarized laser beams propagating perpendicular to $\mb{B}$ (Figs.~\ref{Fig:ExperimentalSetup} and \ref{Fig:ExperimentalGeometry}). The pump lasers are stabilized to the center of the Doppler-broadened $^{85}$Rb $F=2 \rightarrow F'$ hyperfine component of the D2 transition and the center of the $^{87}$Rb $F=2 \rightarrow F'=1$ hyperfine component of the D1 transition, respectively; this optically pumps atoms from these hyperfine levels into the hyperfine levels of interest (which yield the largest optical rotation signals). The pump beams are independently amplitude-modulated by electro-optic modulators at frequencies matching the corresponding Larmor frequencies for the $^{85}$Rb $F=3$ and $^{87}$Rb $F=2$ ground state hyperfine levels, respectively. The duty cycles (20\%) and powers ($\approx 150~\mu W$ for the $^{85}$Rb D2 transition; $\approx 200~\mu W$ for the $^{87}$Rb D1 transition) are chosen to maximize the transverse spin polarization in these levels. During the $\approx 1~{\rm s}$ probe stage, the pump beams are shuttered, and optical rotation of a linearly polarized probe beam is measured with a polarizing beamsplitter and autobalanced photoreceiver. The $\approx 2~{\rm mm}$ diameter probe beam is collinear with the pump beam path. The probe beam is detuned several GHz to the low frequency side of the $^{87}$Rb D2 $F=2 \rightarrow F'$ transition and frequency stabilized using a wavemeter. At this detuning, spin precession of atoms in the $^{85}$Rb $F=3$ and $^{87}$Rb $F=2$ ground state hyperfine levels can be simultaneously measured by detecting optical rotation of the probe light. The time base for the data acquisition is provided by a 10~MHz signal from a GPS-disciplined Rb atomic frequency standard. The time-dependent optical rotation signal measured during the probe phase is Fourier transformed and the resultant peaks are fit to Lorentzians from which the values of $\Omega_{85}$ and $\Omega_{87}$ are extracted (Fig.~\ref{Fig:SampleDataFrequency}). Further details of the experimental apparatus are discussed in Ref.~\cite{Kim13}.

\begin{figure}
\includegraphics[width=3.35 in]{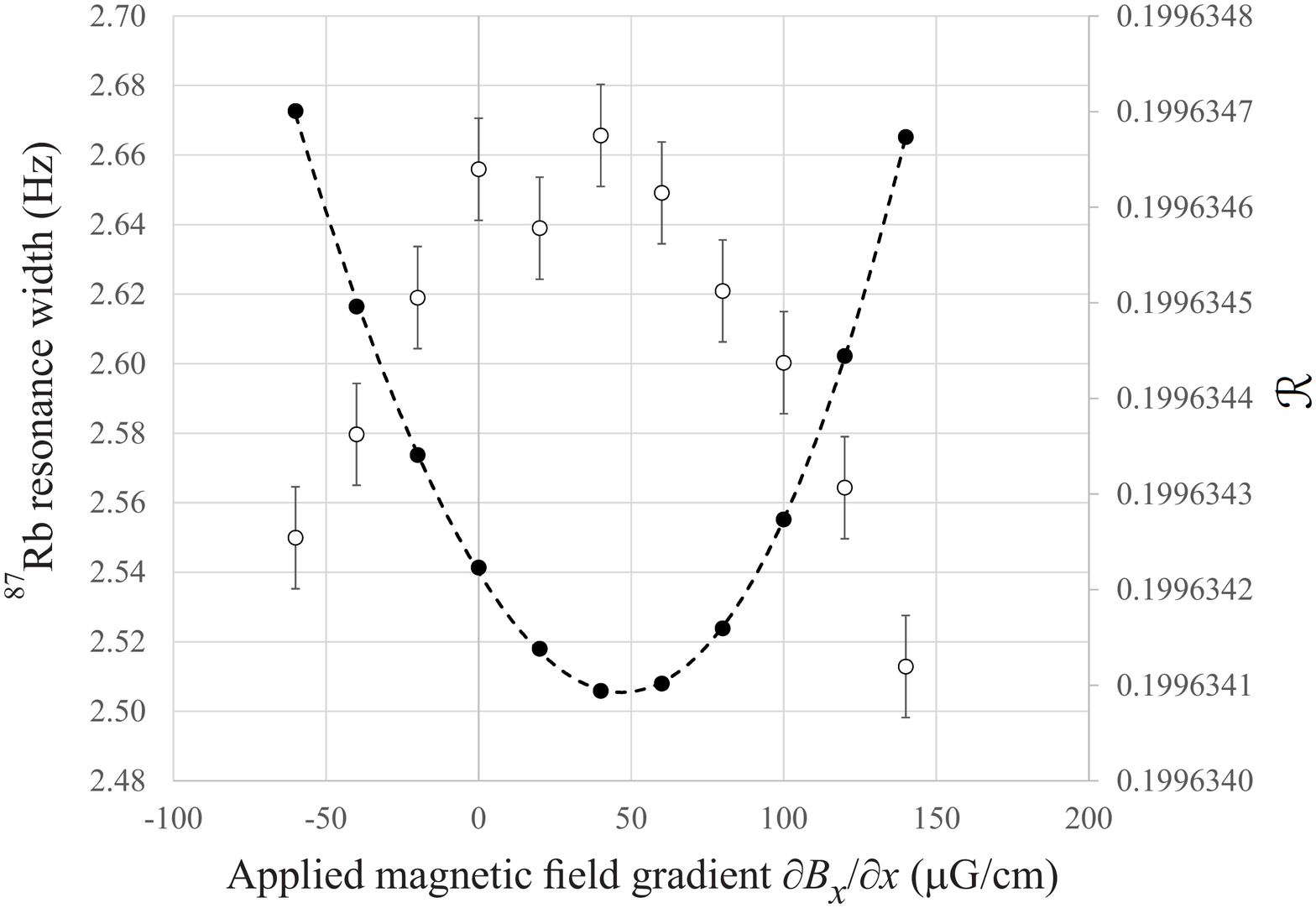}
\caption{Dependence of $\sR$ (open circles) and the width of the Lorentzian fit to the $\Omega_{87}$ peak (filled circles) on applied first-order magnetic field gradient ($\partial B_x/\partial x$). For this data, the alkene-coated vapor cell was used and $B = 19.052729(3)~{\rm mG}$.}
\label{Fig:GradientCompensationExample}
\end{figure}

Each experimental run begins by demagnetizing the innermost magnetic shield and aligning $\mb{B}$ along $\bs{\Omega}_E$ with an accuracy of $\lesssim 1^\circ$ using surveying methods and nonlinear magneto-optical rotation (NMOR) \cite{Bud02,Kim09,Pus06} as discussed in Ref.~\cite{Kim13}; this alignment makes the systematic error associated with the gyro-compass effect quadratic in the misalignment angle between $\mb{B}$ and $\bs{\Omega}_E$ \cite{Ven92}. Magnetic field gradients (that can arise, for example, due to residual magnetization within the innermost shield) cause measurable spin precession frequency shifts \cite{Cat88,She14} that do not exactly cancel in $\sR$ (in spite of motional averaging in the evacuated cell \cite{Pus06b}). Figure~\ref{Fig:GradientCompensationExample} shows both $\sR$ and the width of the Lorentzian fit to the $\Omega_{87}$ peak as a function of the applied gradient $\partial B_x/\partial x$, where $\hat{\mb{x}}$ is orthogonal to the direction of $\mb{B}$ (whose direction is specified to be along $\pm\hat{\mb{z}}$) and the laser beam propagation direction (specified to be along $-\hat{\mb{y}}$), see Fig.~\ref{Fig:ExperimentalGeometry}. Because the fractional effect of gradients on the width is larger than the fractional effect on $\sR$, the gradients can be efficiently compensated by minimizing the widths. It was found that the residual magnetic field gradients would occasionally change when the direction of $\mb{B}$ was reversed, sometimes by several tens of ${\rm \mu G/cm}$, and so it is necessary to recompensate the gradients after every magnetic field reversal. The uncertainty in the compensation of each magnetic field gradient is $\approx 1~{\rm \mu G/cm}$ based on the uncertainty of the fits determining the minimum resonance width, with the important exceptions of $\partial B_x / \partial y$ and $\partial B_y / \partial x$ which have no significant effect on the resonance widths. It is important to recognize that the resonance widths and frequency shifts are affected differently by gradients \cite{Cat88}, and so this method of gradient compensation is not ideal. Specifically, since in our experiment we operate in the regime where the transit time for atoms across the cell ($\approx 0.1~{\rm ms}$) is on the order of or slower than the Larmor period ($\lesssim 0.1~{\rm ms}$), the shift of precession frequencies $\Omega_{85}$ and $\Omega_{87}$ depend more strongly on gradients of the transverse field components $B_x$ and $B_y$, while the widths depend more strongly on gradients of the longitudinal field component $B_z$ \cite{Cat88}. This is because in this regime the frequency shifts are primarily the result of the geometric or Berry's phase effect, while the broadening of the resonance is primarily due to the spatial inhomogeneity of the leading field \cite{Cat88}. While most of the transverse gradients are directly related to longitudinal gradients through Maxwell's equations and can be adequately controlled by measuring the widths, $\partial B_x / \partial y$ and $\partial B_y / \partial x$ are not constrained by this method. As a proxy for a direct measurement of $\partial B_x / \partial y$ and $\partial B_y / \partial x$ under our experimental conditions, we also measure the gradients in near zero-field conditions using the widths of NMOR resonances as described in Ref.~\cite{Pus06b}. Under near zero-field conditions, the transit time across the cell is much faster than the Larmor period, and the NMOR resonance widths are sensitive to both longitudinal and transverse gradients. If the measured gradients change by more than $5~{\rm \mu G/cm}$ either when reversing $\mb{B}$ or when going from near zero field to the value of $\mb{B}$ where spin-precession data are acquired, those data are rejected due to the fact that $\partial B_x / \partial y$ and $\partial B_y / \partial x$ could also have changed by $\gtrsim 5~{\rm \mu G/cm}$ but in an unknown way. Generally we have found that if one gradient component changes by a certain amount when changing $\mb{B}$, several other components also change by similar amounts. We estimate that these procedures for minimizing gradients should lead to systematic offsets no larger than $5~{\rm \mu G/cm}$. Based on the relationship between $\sR$ and the gradients, this yields an overall systematic uncertainty in $\Delta\sR$ of $\lesssim 3 \times 10^{-9}$ (the measured relationship between $\sR$ and the gradients is consistent with calculations based on Refs.~\cite{Cat88,She14}).

In addition, the effect of oscillating magnetic fields on $\sR$ due to the ac Zeeman effect was independently measured. Based on the current noise measured in the coils using a spectrum analyzer, ac-Zeeman-related systematic errors in $\sR$ are negligible in our experiment ($\lesssim 10^{-14}$).

Light shifts due to the probe beam can also affect $\Omega_{85}$ and $\Omega_{87}$ \cite{Mat68,Bul71,Coh72}. The vector light shift can be modeled as a fictitious static magnetic field directed along the light propagation direction \cite{Mat68,Hap72}. Since the probe beam wave vector $\mb{k}\ts{probe}$ is orthogonal to $\mb{B}$ (Fig.~\ref{Fig:ExperimentalGeometry}), the systematic error in $\Delta \sR$ related to the vector light shift is nominally quadratic in the probe beam's ellipticity $\epsilon$. If $\mb{k}\ts{probe}$ deviates from orthogonality to $\mb{B}$ by an angle $\theta$, there can be a component of the fictitious field along $\mb{B}$ leading to a systematic error in $\Delta \sR$ proportional to $\epsilon \theta$. Since the vapor cell walls are somewhat birefringent, we use ellipticity-induced nonlinear magneto-optical rotation with frequency-modulated light (EI FM NMOR) to carry out an \emph{in situ} measurement of the probe beam ellipticity within the cell as described in detail in Ref.~\cite{Kim17}. The probe beam ellipticity prior to entering the cell is adjusted with a quarter-wave plate to minimize the ellipticity within the cell, resulting in $\epsilon \lesssim 2 \times 10^{-4}~{\rm rad}$. Measurement of $\Delta \sR$ as a function of $\epsilon$ allows determination of $\theta$ based on a calculation of the vector light shift as described in Ref.~\cite{Kim13} (noting that due to motional averaging in an antirelaxation-coated cell, the light shift is determined by the cell-volume-averaged intensity of the laser beam \cite{Zhi16}). We find that $\theta \lesssim 4 \times 10^{-4}~{\rm rad}$, in agreement with the estimated sensitivity of the NMOR methods \cite{Pus06} used to minimize $\theta$. At the given probe beam power and detuning, we estimate that the vector light shifts from circularly polarized light propagating along $\mb{z}$ would generate $\Delta \sR \approx 4 \times 10^{-4}$. Thus based on these measurements and estimates, under our experimental conditions the systematic error in $\Delta \sR$ related to vector lights shifts due to the probe beam is $\lesssim 3 \times 10^{-11}$.

\begin{figure}
\includegraphics[width=3.35 in]{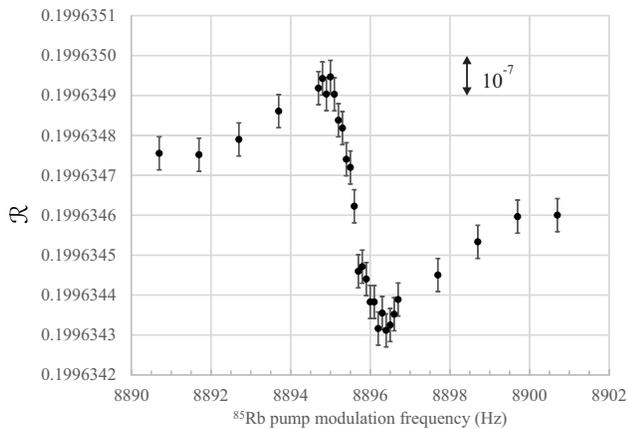}
\caption{The dependence of $\sR$ on the $^{85}$Rb pump laser beam amplitude modulation frequency; $\Omega_{85} \approx 8895.5~{\rm Hz}$ based on fits to the Fourier transform of the probe optical rotation data. For this data, the alkene-coated vapor cell was used and $B = 19.052729(3)~{\rm mG}$.}
\label{Fig:ModulationFrequencyDependence}
\end{figure}

In spite of the fact that the pump beam is blocked during the probe stage in which $\Omega_{85}$ and $\Omega_{87}$ are measured, a number of pump-beam-related systematic effects were discovered during the course of the experiments. Figure~\ref{Fig:ModulationFrequencyDependence} illustrates one of the most prominent effects, a dependence of $\sR$ on the detuning of the $^{85}$Rb pump beam amplitude-modulation frequency from $\Omega_{85}$ (a similar effect is observed for $^{87}$Rb). This effect was first observed in an experiment searching for the permanent electric dipole moment of Hg, and is discussed in detail in Ref.~\cite{Swa13}. Essentially, if the pump modulation frequency is detuned from the spin precession frequency, in the frame rotating with the spins the vector light shift due to the pump beam causes the spins to tip along $\mb{B}$. This is because if the optical pumping is asynchronous with the spin precession, the average direction of the pump wave vector $\mb{k}\ts{pump}$ leads or lags the direction of the spin in the rotating frame. In this case, the spins can precess around the fictitious magnetic field due to the vector light shift from the pump light. This effect is analogous to the action of a rotating transverse magnetic field in magnetic resonance experiments. Spin polarization along $\mb{B}$ generates shifts of $\Omega_{85}$ and $\Omega_{87}$ primarily due to spin-exchange collisions \cite{Hap72,Sch89,Dmi97,Dmi07}. To minimize errors due to this effect, the respective pump modulation frequencies are tuned to within $\lesssim 3~{\rm mHz}$ of $\Omega_{85}$ and $\Omega_{87}$ and the magnetic field is subsequently stabilized using a feedback loop based on measurement of $\Omega_{87}$. This maintains a constant value of $B$ throughout the experiment, limiting shifts of $\Delta\sR$ to $\lesssim 10^{-9}$ due to this asynchronous optical pumping effect. Without active stabilization of $B$, the field magnitude was found to drift by several hundreds of nG during the course of a day-long experimental run.

In addition to the asynchronous optical pumping effect, there is evidence of longitudinal spin polarization generated by scattered pump light. Refraction of light at the coated cell walls leads to scattering of $\approx 10-15\%$ of the light off the back face of the cell (depending on the cell and its position/orientation). If the nominally circularly polarized backscattered light travels preferentially along $\pm\hat{\mb{z}}$ (due to, for example, imperfections in the optical quality of the cell walls or beam misalignment) it can optically pump spin polarization along $\pm\hat{\mb{z}}$. When the vapor cell is initially mounted inside the coil and shield assembly (prior to placing the endcaps and insulation on each shield layer), laser light scattered off the back surface of the cell is observed to reflect at angles of $\approx 5^\circ - 20^\circ$ with respect to the $y$ axis, depending on the cell position and orientation. As in the case of the asynchronous optical pumping effect, longitudinal spin polarization can cause shifts of $\Omega_{85}$ and $\Omega_{87}$ due to spin-exchange collisions. This scattered pump light effect is clearly seen when the quantity
\begin{align}
\Delta \sR_\sigma = \sR({\rm LHC}) - \sR({\rm RHC})
\label{Eq:DeltaR-sigma}
\end{align}
is measured, where $\sR({\rm LHC})$ and $\sR({\rm RHC})$ are the values of $\sR$ for left- and right-circularly polarized pump light, respectively. Due to this effect $\Delta \sR_\sigma \approx \pm 10^{-7}$ in the experiment, and was found to vary at roughly this level when the vapor cell was changed or repositioned. The scattered pump light effect is significantly reduced by reversing the helicity of the pump beams and averaging the result for $\Delta\sR$, which is done automatically throughout the experiment every 80~s. Based on measurements of the pump polarization before and after the cell, we find that the pump helicity reversal is imperfect (due primarily to cell wall birefringence), and so some residual effect likely remains even after averaging. Conservatively assuming that the helicity reversal imperfection occurs entirely at the first cell wall interface encountered by the pump light so that the averaging is minimally effective, the systematic error in $\Delta\sR$ from scattered pump light is estimated to be below $\approx 5 \times 10^{-9}$.

A related potential source of systematic error is a vector light shift from pump light that is re-scattered by other atoms and remains present during the probe phase due to photon diffusion (i.e., radiation trapping, see, e.g., Ref.~\cite{Mat01}). However, the estimated photon diffusion time given the Rb vapor density is only a few ms and the first 20~ms of time-dependent optical rotation during the probe phase is discarded to avoid pump-beam- or shutter-related transient effects. Thus systematic effects related to radiation trapping should be negligible in our experiment.

\begin{figure}
\includegraphics[width=3.35 in]{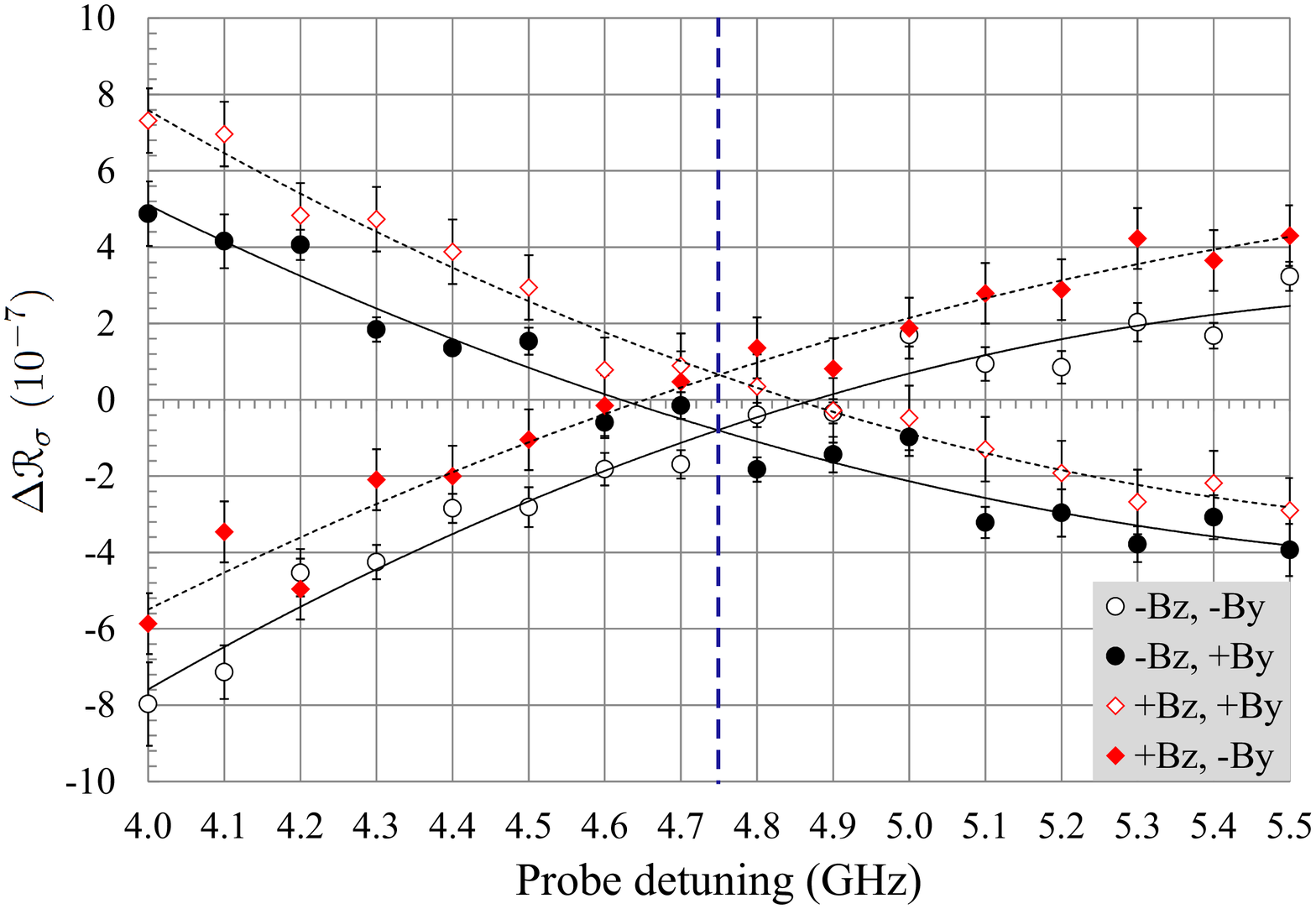}
\caption{Data used to determine the probe laser detuning for which the nonlinear Zeeman effect is compensated by tensor light shifts. The probe detuning is measured relative to the center of the Doppler-broadened $^{87}$Rb D2 $F=2 \rightarrow F'$ transition using a wavemeter. $\mb{B}$ is tilted along $\pm\mb{k}\ts{pump}$ by applying a nonzero $B_y$ component, $\mp B_y$ is along the direction of $\pm \mb{k}\ts{pump}$: filled ($+B_y$) and unfilled ($-B_y$) black circles are for $-B_z$ (the $z$-component of $\mb{B}$ pointing opposite to $\bs{\Omega}_E$), filled ($+B_y$) and unfilled ($-B_y$) red diamonds are for $+B_z$ (the $z$-component of $\mb{B}$ pointing along $\bs{\Omega}_E$). The detuning for which $\Delta R_\sigma$ is equal for both $+B_y$ and $-B_y$ is the compensation point, indicated by the blue dashed vertical line. The compensation point for $\pm B_z$ is the same within uncertainty. $\Delta R_\sigma$ is offset from zero at the compensation point due to scattered pump light along $\hat{\mb{z}}$. For this data, the alkene-coated vapor cell was used and $B = 19.052729(3)~{\rm mG}$.}
\label{Fig:TensorShiftCancellation}
\end{figure}

A significant second-order systematic effect related to tensor shifts was also observed. There are two principal independent causes of tensor shifts in our experiment: tensor light shifts due to the probe beam and the nonlinear Zeeman effect \cite{Aco06,Jen09,Cha10}. The first-order effect of tensor shifts is merely to broaden the spin precession resonance, and under our experimental conditions the tensor-shift broadening is generally negligible compared to other effects. However, tensor shifts in the presence of longitudinal spin polarization, as can occur if there is a nonzero projection of $\mb{k}\ts{pump}$ along $\mb{B}$, create an asymmetry in the spin precession resonance lineshape causing apparent shifts of $\Omega_{85}$ and $\Omega_{87}$. Fortunately, the tensor light shift can be used to cancel the nonlinear Zeeman shift, as demonstrated in Ref.~\cite{Jen09}. In order to carry out this compensation, the linear polarization of the probe beam $\hat{\bs{\varepsilon}}$ is adjusted to be along the $x$-axis (orthogonal to $\mb{B}$, see Fig.~\ref{Fig:ExperimentalGeometry}), in which case the effect of the tensor light shift on $\Delta \sR$ has the opposite sign as that of the nonlinear Zeeman effect. This systematic effect can be made larger by intentionally tilting $\mb{B}$ along $\mb{k}\ts{pump}$ in order to increase longitudinal spin polarization. This is done by applying a nonzero component of $\mb{B}$ along $\pm\hat{\mb{y}}$ (and respectively reducing $B_z$) in order to tilt $\mb{B}$ by $\approx \pm 7^\circ$ along $\mb{k}\ts{pump}$. Since the sign of the tensor shift systematic reverses with longitudinal spin polarization, by measuring $\Delta \sR_\sigma$ [Eq.~\eqref{Eq:DeltaR-sigma}] as a function of probe detuning (Fig.~\ref{Fig:TensorShiftCancellation}), the detuning for which cancellation between the tensor light shift and nonlinear Zeeman effect occurs can be determined. The probe detuning is stabilized to this value and $\mb{B}$ is tilted back to its original direction along $\hat{\mb{z}}$, orthogonal to $\mb{k}\ts{pump}$, as in Fig.~\ref{Fig:ExperimentalGeometry}. The wave vector $\mb{k}\ts{pump}$ is carefully aligned to $\mb{k}\ts{probe}$ using quadrant photodiodes before and after the cell (separated by $\approx 100~{\rm cm}$), ensuring that they are aligned with one another to within $\approx 10^{-4}~{\rm rad}$. Data used to determine the probe detuning for the tensor shift compensation are shown in Fig.~\ref{Fig:TensorShiftCancellation}. Reversing the projection of $\mb{B}$ along $\mb{k}\ts{pump}$ by applying $\pm B_y$ reverses the longitudinal spin polarization; the intersection of the two $\Delta R_\sigma$ curves for $\pm B_y$ shows the probe detuning where the tensor shift compensation occurs --- note that the compensation point is the same for $\pm B_z$ as expected. The offsets of $\Delta R_\sigma$ from zero at the compensation point are due to the scattered light effect discussed above, which does not change appreciably for field tilts of $\approx \pm 7^\circ$. Based on the product of the uncertainty in the compensation point due to the statistical errors of the fits and the uncertainty in pump beam alignment, the systematic error in $\Delta \sR$ due to tensor shifts is $\lesssim 2 \times 10^{-10}$.

Other physical effects that could, in principle, cause systematic errors in the determination of $\Delta \sR$, such as spin-exchange collisions between the Rb isotopes with precessing transverse spin polarization and frequency shifts due to the nuclear magnetic moments, were considered in Ref.~\cite{Kim13}, and estimated upper limits on such effects are listed in Table~\ref{Table:systematic-errors} along with those discussed in the present work.

\begin{table}
\caption{Estimated upper limits on the contributions of various sources of systematic errors to $\Delta \sR$. Those marked with $^*$ are discussed in Ref.~\cite{Kim13}.}
\medskip \begin{tabular}{lc} \hline \hline
Description & Effect on $\Delta \sR$ \\
\hline
\rule{0ex}{3.6ex} Scattered pump light along $\mb{B}$ & $5\times 10^{-9}$  \\
\rule{0ex}{3.6ex} Magnetic field gradients & $3 \times 10^{-9}$ \\
\rule{0ex}{3.6ex} Excess noise for $-\mb{B}$ & $3 \times 10^{-9}$  \\
\rule{0ex}{3.6ex} Asynchronous optical pumping & $10^{-9}$  \\
\rule{0ex}{3.6ex} Tensor shifts + polarization along $\mb{B}$ & $2\times 10^{-10}$  \\
\rule{0ex}{3.6ex} Vector light shifts from probe beam $\epsilon$ & $3 \times 10^{-11}$  \\
\rule{0ex}{3.6ex} Gyro-compass effect & $10^{-13}$ \\
\rule{0ex}{3.6ex} ac Zeeman effect & $10^{-14}$  \\
\rule{0ex}{3.6ex} Wall collisions$^*$ & $10^{-16}$  \\
\rule{0ex}{3.6ex} Nuclear magnetic moments$^*$ & $10^{-16}$  \\
\rule{0ex}{3.6ex} Transverse spin-exchange collisions$^*$ & $2 \times 10^{-18}$  \\
\hline \hline
\end{tabular}
\label{Table:systematic-errors}
\end{table}

Taking into account the various systematic errors discussed above, the experimental procedure involves a number of steps which are summarized as follows. As noted above, experimental runs begin by de-gaussing the magnetic shields and aligning the magnetic shield axis along $\bs{\Omega}_E$. The pump and probe beams are aligned to be collinear with one another using the quadrant photodiodes positioned before and after the magnetic shields. Transverse magnetic fields and magnetic field gradients are measured using NMOR techniques at near-zero magnetic fields \cite{Bud02,Kim09,Pus06}. Next, EI FM NMOR techniques \cite{Kim17} are used to zero the \emph{in situ} ellipticity of the probe light within the vapor cell. Then the working value of the magnetic field $\mb{B}$ is applied and a transverse component along $+\mb{y}$ is added to tilt the field by $+7^\circ$. The gradients are compensated by measuring the resonance widths as a function of applied gradients (see Fig.~\ref{Fig:GradientCompensationExample}). Then the probe beam detuning is scanned while carrying out the pump/probe measurement of $\Omega_{85}$ and $\Omega_{87}$ to determine the tensor-shift-related change in $\Delta R_\sigma$ (Fig.~\ref{Fig:TensorShiftCancellation}). Next a transverse field component along $-\mb{y}$ is added, the gradients are re-compensated, and the probe beam detuning is scanned again to re-measure the tensor-shift-related change in $\Delta R_\sigma$ and find the probe beam detuning for which the tensor light shifts compensate the nonlinear Zeeman shifts. The probe beam frequency is then locked to this compensation value. The transverse fields are now compensated by finding the minimum value of $\Omega_{87}$ as a function of applied fields along $x$ and $y$. The gradients are again re-compensated by measuring the resonance widths. If the change in gradients from the near-zero field values is greater than $5~{\rm \mu G /cm}$, the entire process is repeated. Once the gradients are stable and well-compensated, the data for measuring $\sR_+$ are acquired: 40 two-second pump/probe measurements of $\Omega_{85}$ and $\Omega_{87}$ are acquired with LHC-polarized pump light, then 40 two-second pump/probe measurements of $\Omega_{85}$ and $\Omega_{87}$ are acquired with RHC-polarized pump light. This is repeated 16 times for a total of 1280 individual measurements of $\sR_+$. Then the field is reversed, the tensor shifts and gradients are re-measured and compensated, and the latter steps are repeated to measure $\sR_-$.

\begin{figure}
\includegraphics[width=3.35 in]{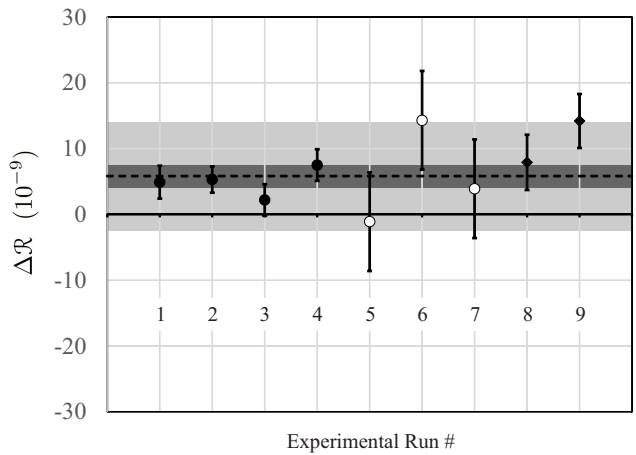}
\caption{Value of $\Delta \sR$ extracted from different runs. Each run consisted of 1280 individual acquisitions for both $\pm \mb{B}$. The first four data points representing experimental runs 1-4 (black filled circles) were taken with $B = 19.052729(3)~{\rm mG}$ using an alkene-coated vapor cell. The next three data points (black unfilled circles) representing experimental runs 5-7 were taken with $B = 28.579094(3)~{\rm mG}$ using an alkene-coated vapor cell. The last two data points (black filled diamonds) representing experimental runs 8 and 9 were taken with $B = 19.052729(3)~{\rm mG}$ using a paraffin-coated vapor cell. The dashed line represents the weighted average of the results, the thin dark gray band represents the statistical uncertainty of the weighted average of the results, and the thicker light gray band represents the overall systematic uncertainty as determined by adding in quadrature the various estimates of systematic errors listed in Table~\ref{Table:systematic-errors}.}
\label{Fig:GDMresults}
\end{figure}

In order to compensate and control various systematic effects, the chosen values of many of the experimental parameters are interconnected, which limits the ability to independently vary parameters. To check for unknown systematic effects, data were taken at two different magnetic field magnitudes: $B = 19.052729(3)~{\rm mG}$ (denoted the low field value) and $B = 28.579094(3)~{\rm mG}$ (denoted the high field), where the field magnitudes were determined by measurement of $\Omega_{87}$. The probe beam power and detuning were adjusted accordingly for each field magnitude to compensate the tensor shift systematic effect as discussed above and as shown in Fig.~\ref{Fig:TensorShiftCancellation}. Experimental runs 1-4, 8 and 9 (black filled circles and black filled diamonds in Fig.~\ref{Fig:GDMresults}) were taken at the low field magnitude, while experimental runs 5-7 (black unfilled circles in Fig.~\ref{Fig:GDMresults}) were taken at the high field magnitude.  At the high field magnitude, excessive non-statistical point-to-point fluctuations of the data were observed for the $-\mb{B}$ data in particular, leading to considerably larger error bars for the high-field results. This may be due to excess current noise from the voltage supply for negative applied voltages at the higher field. The excess noise is also manifested in a larger number of poorer quality fits to determine $\Omega_{85}$ and $\Omega_{87}$ for the $-\mb{B}$ data at high field, which were systematically biased to result in a smaller value of $\sR_{-}$ and thus a larger value of $\Delta \sR$. By excluding individual data points from the average if their fit uncertainty for $\Omega_{85}$ and $\Omega_{87}$ exceeded 1~mHz, the mean value of $\Delta \sR$ was found to shift by up to $3 \times 10^{-9}$. Thus we estimate that the contribution to the systematic error in $\Delta \sR$ due to excess noise for $-\mb{B}$ data is smaller than $\approx 3 \times 10^{-9}$.

Another source of spin-precession-frequency shifts is the interaction of atoms with the cell walls. Wall collisions in vapor cells that have shapes with quadrupolar anisotropy have been shown to cause tensor shifts (see, for example, Ref.~\cite{Pec16} and references therein) due to the interaction of the atomic electric quadrupole moment with electric field gradients at the cell wall surfaces (asymmetric electric fields coupling to atomic spins through the tensor polarizability may also play a significant role in this effect \cite{Pec16}). Wall-induced effects are minimized in our experiment in two different ways: first, we use a spherical cell which has small quadrupolar anisotropy (only due to the presence of a stem containing the Rb metal); second, we directly compensate tensor shifts by adjusting the tensor light shift as described above (Fig.~\ref{Fig:TensorShiftCancellation}), which should zero any tensor shifts caused by wall collisions. Effects due to wall collisions were estimated to be entirely negligible for our experiment in Ref.~\cite{Kim13}, but as a precaution data were taken with two different cells with different coatings (alkene \cite{Bal10} and paraffin \cite{Ale02}). The stems of the cells were also oriented differently so as to change the quadrupolar shape anisotropy between the experimental runs. No evidence of a systematic shift between the data for the two cells was found (Fig.~\ref{Fig:GDMresults}).

Based on these measurements, we find that
\begin{align}
\Delta \sR = 5.8 \pm 1.7\ts{(stat)} \pm 6.6\ts{(sys)} \times 10^{-9}~,
\label{Eq:Delta-R-result}
\end{align}
where the mean and statistical uncertainty is based on the weighted average of the results shown in Fig.~\ref{Fig:GDMresults} and the systematic uncertainty is determined from adding the estimated systematic errors from Table~\ref{Table:systematic-errors} in quadrature. Combining the statistical and systematic uncertainties in quadrature yields an upper limit on $\Delta \sR$:
\begin{align}
\Delta \sR \leq 1.5 \times 10^{-8}~~({\rm 90\%~confidence})~.
\label{Eq:Delta-R-limit}
\end{align}
From the upper limit on $\Delta \sR$, we derive from Eq.~\eqref{Eq:DeltaR} an upper limit on the proton gyro-gravitational ratio:
\begin{align}
\chi_p \leq 5.6 \times 10^{-33}~{\rm g \cdot cm}~~({\rm 90\%~confidence})~,
\end{align}
which in turn, based on Eq.~\eqref{Eq:k}, gives the upper limit on proton GDM parameter $k_p$ listed in Table~\ref{Table:k-limits}, over three orders of magnitude more stringent than the existing constraint from Ref.~\cite{You96}. This implies through Eq.~\eqref{Eq:GDM-Hamiltonian} that the spin-dependent part of the gravitational energy of the proton is $\leq 3.4 \times 10^{-18}~{\rm eV}$.

\begin{figure}
\includegraphics[width=4.2 in]{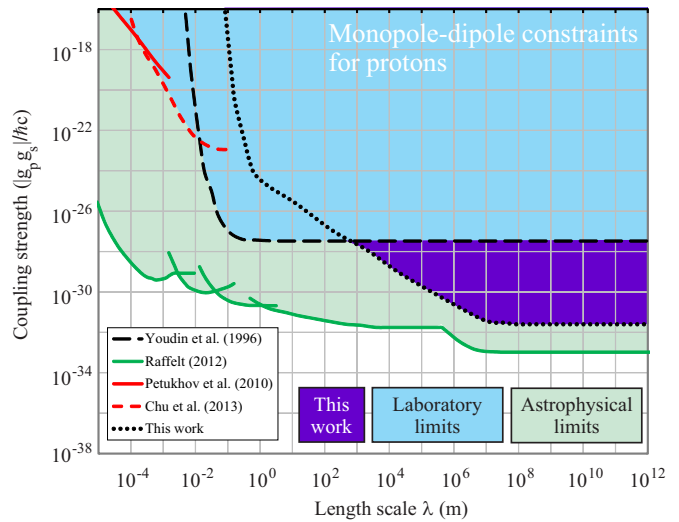}
\caption{Constraints on monopole-dipole (scalar-pseudoscalar) proton couplings, $\left| {\rm g}_p {\rm g}_s \right|/\hbar c$ as a function of the range $\lambda$ of the interaction [${\rm g}_p$ and ${\rm g}_s$ are the pseudoscalar and scalar coupling constants, respectively, see Eq.~\eqref{Eq:monopole-dipole-potential}]. Parameter space excluded by previous laboratory experiments is shaded light blue; the dashed black line shows results from Ref.~\cite{You96}, the solid red line is from Ref.~\cite{Pet10}, and the dashed red line is from Ref.~\cite{Chu13}. Astrophysical constraints (excluded parameter space shaded light green) are from the analysis of \citet{Raf12}. The dotted black line and purple shading represent the constraints derived from the present measurement.}
\label{Fig:MonopoleDipoleConstraints}
\end{figure}

If the results of our experiment are interpreted as a constraint on long-range monopole-dipole couplings of the proton based on Eqs.~\eqref{Eq:monopole-dipole-potential} and \eqref{Eq:k}, they exclude the parameter space shown in Fig.~\ref{Fig:MonopoleDipoleConstraints} outlined with the dotted black line and shaded purple. In the long-range limit where $\lambda \rightarrow \infty$, we find an upper limit on the monopole-dipole coupling constant for the proton of
\begin{align}
\frac{\left|{\rm g}_p {\rm g}_s \right|}{\hbar c} \leq 2.5 \times 10^{-32}~~({\rm 90\%~confidence})~.
\end{align}
The astrophysical constraints on $\left| {\rm g}_p {\rm g}_s \right|/\hbar c$ (see Ref.~\cite{Raf12}) are more stringent than the constraints obtained in our experiment, although in the case of the astrophysical constraints there is both a degree of model specificity \cite{Mas05} and some degree of uncertainty regarding the accuracy of stellar models. It is also possible that a so-called ``chameleon mechanism'' could screen such interactions in regions of space with high mass density, invalidating astrophysical bounds on new interactions \cite{Jai06}. Furthermore, if there exist both new spin-0 and spin-1 bosons that interact with each other, the astrophysical bounds can be significantly weakened \cite{Red08}. Thus direct laboratory measurements play a crucial, comparatively less ambiguous role in determining the existence of exotic spin-dependent interactions.

In conclusion, we have searched for a long-range monopole-dipole coupling between the mass of the Earth and Rb nuclear spins. Our measurement constrains spin-gravity couplings and long-range monopole-dipole couplings of the proton over three orders-of-magnitude more stringently than previous laboratory limits \cite{You96}. We note that there are several promising new ideas that could lead to improved constraints on spin-gravity interactions, including new nuclear-spin comagnetometers \cite{She14,Led12} and an experiment based on a precessing ferromagnetic needle \cite{Kim16-needle}. Our measurement should provide a more precise determination of the ratio of the $^{87}$Rb and $^{85}$Rb gyromagnetic ratios ($\gamma_{87}/\gamma_{85}$) as compared to the present best measurement which is at the parts-per-million level \cite{Whi68,Cha11}, although systematic errors may contribute somewhat differently to $\gamma_{87}/\gamma_{85}$ and will be evaluated in a future work. Furthermore, our measurement should improve constraints on long-range velocity- and spin-dependent interactions between protons and polarized electrons in the Earth as discussed in Refs.~\cite{Hun13,Hun14}.




\acknowledgments

We are deeply indebted to generations of undergraduate students who spent countless hours over many years working on earlier iterations of this experiment, especially Rene Jacome, Ian Lacey, Jerlyn Swiatlowski, Eric Bahr, Srikanth Guttikonda, Khoa Nguyen, Rodrigo Peregrina-Ramirez, Lok Fai Chan, Cesar Rios, Caitlin Montcrieffe, Claudio Sanchez, and Swecha Thulasi. The authors are also sincerely grateful to Dmitry Budker, Blayne Heckel, Michael Romalis, Jennie Guzman, Tuan Nguyen, Larry Hunter, Max Zolotorev, and Eugene Commins for invaluable discussions and to Mohammad Ali and Li Wang for technical work on parts of the apparatus. The magnetic shield system was designed by Valeriy Yashchuk and the coil system for magnetic field control was co-designed with Valentin Dutertre. The antirelaxation-coated vapor cells were manufactured by Mikhail Balabas. This work was supported by the National Science Foundation under grants PHY-0652824, PHY-0969666, and PHY-1307507.  The findings expressed in this material are those of the authors and do not necessarily reflect those of the NSF.

\bibliography{GDM-result-bib}

\end{document}